\documentclass[showpacs,floatfix,prl,twocolumn, superscriptaddress]{revtex4}

\usepackage{float}
\usepackage{graphicx}

\newcommand{\bk}{{\bf k}}

\newcommand{\beq}{\begin{eqnarray}}
\newcommand{\eeq}{\end{eqnarray}}
\newcommand{\beqq}{\begin{eqnarray*}}
\newcommand{\eeqq}{\end{eqnarray*}}

\begin{document}

\title{Odd-parity topological superconductor with nematic order: application to Cu$_{x}$Bi$_2$Se$_3$}

\author{Liang Fu}
\affiliation{Department of Physics, Massachusetts Institute of Technology, Cambridge, MA 02139}

\begin{abstract}
Cu$_{x}$Bi$_2$Se$_3$ was recently proposed as a promising candidate for time-reversal-invariant topological superconductors\cite{FuBerg}. In this work, we argue that the unusual anisotropy of the Knight shift observed by Zheng {\it et al}\cite{zheng}, taken together with specific heat measurements, provides strong support for an unconventional odd-parity pairing in the two-dimensional $E_u$ representation of the $D_{3d}$ crystal point group\cite{FuBerg}, which spontaneously breaks the three-fold rotational symmetry of the crystal, leading to a subsidiary nematic order. 
We predict that the spin-orbit interaction associated with hexagonal warping plays a crucial role in pinning the two-component order parameter and 
makes the superconducting state fully-gapped, leading to a  topological superconductor. Experimental signatures of the $E_u$ pairing related to the nematic order are discussed. 
\end{abstract}

\maketitle

Time-reversal-invariant (T-invariant) topological superconductors in two and three dimensions are a new class of unconventional superconductors which 
exhibit a full superconducting gap in the bulk and gapless helical quasi-particles on the boundary\cite{Ludwig, Raghu, Roy}. Because these quasi-particles  do not possess conserved charge or spin quantum numbers, they cannot be distinguished from their anti-particles and hence are regarded as itinerant Majorana fermions. %, with exotic properties\cite{Chung, Ryu}. 

There is currently intensive effort
 in finding T-invariant topological superconductors in real materials\cite{Qi, Lee, Nagaosa, Schamlian, Zhang}. Recent theoretical works\cite{FuBerg, Sato} have established that the key requirement for topological superconductivity in inversion-symmetric systems is odd-parity pairing symmetry. 
Only a few odd-parity superconductors are known to date. Two   
prime examples are Sr$_2$RuO$_4$ and UPt$_3$. However, both materials seem to have nodes and/or spontaneously break time-reversal symmetry, hence do not qualify as T-invariant topological superconductors.  

Recently, the doped topological insulator Cu$_x$Bi$_2$Se$_3$, which is superconducting with a maximum $T_c$ of 3.8K\cite{Cava}, was proposed as a candidate topological superconductor with odd-parity pairing\cite{FuBerg}. Since then this material has been intensively studied. Specific heat measurements down to $0.3$K found a full superconducting gap\cite{KrienerPRL}. 
The upper critical field appears to exceed the Pauli limit, which is interpreted as consistent with triplet pairing\cite{deVisser}.  
Much interest is sparked by the observation of a zero-bias conductance peak in a point-contact spectroscopy experiment on Cu$_{0.3}$Bi$_2$Se$_3$\cite{pointcontact}, which is attributed to the putative Majorana fermion surface states  from topological superconductivity. However, a later scanning tunneling spectroscopy measurement on Cu$_{0.2}$Bi$_2$Se$_3$ found a full gap in the tunneling spectrum  at very lower temperature, without any sign of in-gap states\cite{Stroscio}. The discrepancy between these two {\it surface sensitive} experiments has led to considerable debate and controversy about the nature of superconductivity in Cu$_x$Bi$_2$Se$_3$\cite{pt2, ARPES, Tanaka, ep, sau}. In view of the current status, direct probes of the pairing symmetry in the {\it bulk} are much needed. 
  
In a very recent nuclear magnetic resonance (NMR) study of Cu$_{0.3}$Bi$_2$Se$_3$, Zheng's group discovered an unusual  anisotropy in the Knight shift as a  small applied  field is rotated within the $ab$-plane\cite{zheng}. The Knight shift is isotropic above $T_c$, and decreases in the superconducting state. 
Remarkably, the change in the Knight shift is largest when the field is parallel to a particular crystal axis.  
This  in-plane uniaxial anisotropy does not conform with the three-fold rotational symmetry of the crystal, and thus provides a direct evidence 
of spontaneous crystal symmetry breaking associated with unconventional superconductivity in Cu$_x$Bi$_2$Se$_3$.       

In this work, we identify the pairing symmetry of Cu$_x$Bi$_2$Se$_3$ from the existing NMR and specific heat measurements,  theoretically establish a novel fully-gapped topological superconductor phase, and predict  experimental signatures for further study. Our main finding is that among all possible pairing symmetries, the odd-parity pairing in the two-dimensional (2D) $E_{u}$ representation, 
first introduced in Ref.\cite{FuBerg}, is the only one compatible with the rotational symmetry breaking observed in NMR measurement\cite{zheng} {\it and} the full superconducting gap found in specific heat measurement\cite{KrienerPRL}. 
Since this $E_u$ pairing generates a subsidiary nematic order, we call the resulting state a ``nematic superconductor''. 

The  fully-gapped nature of the $E_u$  superconducting state  found here is remarkable, considering that all previous works found  
nodes in this state\cite{FuBerg, pointcontact, yip}. Moreover, a full gap is required for topological superconductivity. 
In contrast to previous works that are based on a rotationally invariant Dirac fermion model for the bulk band structure of Cu$_x$Bi$_2$Se$_3$,    
we find that crystalline anisotropy plays an indispensable role in the  odd-parity $E_u$ state. 
We show by general argument and model study that the spin-orbit interaction associated with hexagonal warping\cite{Fu} pins the direction of the two-component $E_u$ order parameter to a two-fold axis of the crystal, consistent with the Knight shift anisotropy, and makes the superconductor fully-gapped.  
Such a nematic superconductor constitutes a new phase of odd-parity pairing.

{\bf Pairing Symmetry}: It was recognized at the outset that strong spin-orbit coupling must be taken into consideration in discussing the pairing symmetry of Cu$_x$Bi$_2$Se$_3$\cite{FuBerg}. Indeed, the importance of spin-orbit coupling becomes manifest in the Knight shift measurement of electron's spin susceptibility. If spin-orbit coupling were absent, the Knight shift would be fully isotropic for spin singlet as well as triplet pairing, because the triplet $d$-vector would be free to rotate with the applied magnetic field. In contrast, in the presence of spin-orbit coupling, the notion of spin-singlet or triplet pairing is, strictly speaking, not well-defined. Instead, pairing symmetries are classified according to the representations of the crystalline symmetry group $D_{3d}$\cite{FuBerg}, which acts on spatial coordinates and electron's spin simultaneously. 
The consequence is that the spin structure of the superconducting order parameter is locked to crystal axis, resulting in an anisotropic spin susceptibility.

Among the the six irreducible representations of $D_{3d}$ ($A_{1g}$, $A_{1u}$, $A_{2u}$,  $A_{2g}$, $E_u$ and $E_g$), only the $E_u$ and $E_g$ representations are multi-dimensional and hence potentially compatible with the spontaneous rotational symmetry breaking observed in the Knight-shift measurement. 
In order to determine which one of the two is the pairing symmetry of Cu$_x$Bi$_2$Se$_3$, we first consider Ginzburg-Landau theory for the $E_u$ and $E_g$ superconducting states. 
The $D_{3d}$ point group symmetry dictates that up to the fourth order, the Landau free energy in both cases must take the form
\beq
F &=& r (|\Psi_1|^2 + |\Psi_2|^2) + u_1 (|\Psi_1|^2 + |\Psi_2|^2)^2  \nonumber \\
&+& u_2  | \Psi_1^2 + \Psi_2^2|^2 \label{f}
\eeq 
where $r \propto (T-T_c)$. Here $\Psi = (\Psi_1, \Psi_2)$ is the two-component order parameter, which transforms like a vector under the three-fold rotation. 
The same form of the free energy also applies to other crystal systems\cite{chubukov, mineev}.  
Importantly, the nature of the superconducting state below $T_c$ depends on the sign of $u_2$. For $u_2>0$, a T-breaking chiral state with a complex order parameter $\Psi \propto (\frac{1}{\sqrt{2}}, \frac{i}{\sqrt{2}})$ arises, which is {\it isotropic} within the $ab$-plane. For $u_2<0$, a T-invariant state with a real order parameter $\Psi \propto (\cos \theta, \sin \theta)$ arises. 
This superconducting state spontaneously breaks the rotational symmetry, and possesses a subsidiary nematic order parameter $Q$: 
\beq
Q =(  | \Psi_1|^2 - | \Psi_2|^2, \;   \Psi_1^* \Psi_2 + \Psi_2^* \Psi_1 ). \label{Q}
\eeq 
The two components of $Q$ transform as $x^2 - y^2$ and $xy$ respectively. 
Such a nematic superconductor with uniaxial anisotropy is consistent with with the Knight shift measurement, whereas the isotropic chiral state is not. 
%The free energy (\ref{f}) with $u_2<0$ leads to a $U(1)$ degeneracy between all values of $\theta$. Higher order terms to be included later in the free energy will remove this extensive degeneracy and pin the nematic director along certain crystalline axes. 

We now show that the nematic state with $E_g$ pairing and the one with $E_u$ pairing can be experimentally distinguished by their qualitatively different gap structures, 
because of the difference in the parity of the order parameter: $E_g$ is even-parity and $E_u$ is odd-parity. To analyze the gap structure, it is convenient to express the pair potential $\Delta(\bk)$ in the band basis. Since the superconducting gap is much smaller than the Fermi energy in Cu$_x$Bi$_2$Se$_3$, it suffices to  consider only the 
bands at the Fermi energy. Due to the presence of both time reversal and inversion symmetry, the energy bands are twofold degenerate at every $\bf k$, which we label by a ``pseudospin'' index $\alpha$. Because of spin-orbit coupling, $\alpha=1,2$ does not correspond to electron's spin. 
The pair potential thus reduces to a $2\times 2$ matrix  over the Fermi surface: the gap function $\Delta_{\alpha \alpha'}(\bk)$.   
%\beq
%\Delta_{i s, j s'} (\bk ) c^\dagger_{\bk, i s} \epsilon_{ ss'} c^\dagger_{-\bk, j s'} = %\tilde{\Delta}_{\alpha \beta} (\bk ) c^\dagger_{\bk \alpha} c^\dagger_{-\bk \beta} \equiv 
%\Delta_{\alpha \beta} (\bk ) c^\dagger_{\bk \alpha} \epsilon_{\beta \beta'}c^\dagger_{-\bk \beta'} + ... , 
%\eeq
%where $i, j$ labels orbitals within a unit cell, $s, s'= \uparrow, \downarrow$ labels electron's spin, and 

Depending on the parity of the order parameter, the gap function of a T-invariant superconductor takes two different forms:  
\beq
\Delta^e(\bk) &=& \Delta(\bk) \cdot I, \; \mbox{where } \Delta(\bk) = \Delta(-\bk), \label{even}  \\
\Delta^o(\bk) &=& {\vec d}(\bk) \cdot {\vec \sigma}, \; \mbox{where } {\vec d}(\bk) =- {\vec d}(-\bk).  \label{odd}
\eeq  
The even-parity gap function $\Delta^e(\bk)$ is a real {\it scalar}, while the odd-parity gap function $\Delta^o(\bk)$ is parameterized by a real {\it vector} field ${\vec d}(\bk)$, the $d$-vector.  The  superconducting gaps $\delta(\bk)$ in the two cases are given by 
$| \Delta(\bk) |$ and $ |{\vec d}(\bk)|$ respectively.  
  
The scalar nature of even-parity gap function (\ref{even}) dictates that the  T-invariant $E_g$  state of Cu$_x$Bi$_2$Se$_3$ must have line nodes. 
To see this,  let us recall that for  any non-s-wave pairing, the gap function integrated over the Fermi surface must be zero:
\beq
\int_{\bk \in {\rm FS} } d \bk \;  \Delta(\bk) = 0. \label{av}
\eeq    
As shown by angle-resolved photoemission spectroscopy experiments\cite{ARPES, arpes2}, Cu$_x$Bi$_2$Se$_3$ has a connected Fermi surface enclosing  $\bk=0$.  
It then follows from Eq.(\ref{av}) that $\Delta(\bk)$ must change sign somewhere on such a  Fermi surface, resulting in unavoidable line nodes. 
As an explicit example,  the $E_g$ gap function $\Delta(\bk) \propto k_z k_x, k_z k_y$ considered in Ref.\cite{yip} has lines of nodes on the $k_z=0$ and $k_x, k_y=0$ planes. The existence of line nodes conflicts with the specific heat measurement\cite{KrienerPRL}. 
This seems sufficient to rule out the $E_g$ pairing  in Cu$_x$Bi$_2$Se$_3$. In contrast, 
we will show below  that the $E_u$ states generically have a full superconducting gap.

{\bf Superconducting gap}:  For the sake of concreteness, we first derive the superconducting gap of the $E_u$ state within a two-orbital model for Cu$_x$Bi$_2$Se$_3$.  Later, we will show that the presence or absence of nodes is a robust property that depends only on symmetry, not microscopic details. 

The band structure of Cu$_x$Bi$_2$Se$_3$ at low energy is described by a $k\cdot p$ Hamiltonian at $\Gamma$, 
which to {\it first order} in $k$ takes the following form\cite{FuBerg}
\beq
H_0 = \sum_\bk c_\bk^\dagger \left[ v ( k_x s_y - k_y s_x) \sigma_z  + v_z k_z \sigma_y + m \sigma_x - \mu \right] c_\bk, \nonumber 
%\\ \label{h0} 
\eeq     
where $c^\dagger=(c^\dagger_{1\uparrow}, c^\dagger_{1\downarrow}, c^\dagger_{2\uparrow}, c^\dagger_{2\downarrow})$  
consists of two orbitals  hereafter denoted as $1$ and $2$, in addition to electron's spin. Here $\sigma$ and $s$ are two sets of Pauli matrices associated with orbital and spin respectively. 
It is worth pointing out that spin-orbit coupling in time-reversal and inversion symmetric systems necessarily involves more than one orbitals, as shown in the two-orbital Hamiltonian here. The physical origin of $H_0$ is elucidated in Ref.\cite{Hsiehfu}. The chemical potential $\mu$ lies in the conduction band due to Cu-doping.  

In this two-orbital model, the $E_u$ pairing arises when electrons in the two orbitals within a unit cell pair up to form a spin triplet, with zero total spin along an in-plane direction ${\bf n}=(n_x, n_y)$.  The corresponding pair potential, $V_{\bf n} = n_x V_x + n_y V_y$, is a superposition of two independent basis functions given in Ref.\cite{FuBerg} (therein called ``$\Delta_4$ pairing''):  
\beq
V_x &=& i \Delta_0 ( c^\dagger_{1\uparrow}  c^\dagger_{2\uparrow} -  c^\dagger_{1\downarrow}  c^\dagger_{2\downarrow})  \nonumber \\
V_y &=&    \Delta_0 (c^\dagger_{1\uparrow}  c^\dagger_{2\uparrow} + c^\dagger_{1\downarrow}  c^\dagger_{2\downarrow}). 
 \label{pp}
\eeq 
$V_{\bf n}$ is T-invariant and rotational symmetry breaking.  
$\bf n$ should be regarded as a nematic director (a headless vector), because the superconducting order parameter $V_{\bf n}$ and $V_{-\bf n}$ only differ by sign and correspond to the same physical state. 
 
We can  directly obtain the superconducting gap $\delta_{\bf n}(\bk)$ by diagonalizing the BCS mean-field Hamiltonian $H_{sc}=H_0 + V_{\bf n}$. 
Alternatively, we can derive the gap function $\Delta(\bk)$ by rewriting $V_{\bf n}$, defined by (\ref{pp}) in spin and orbital basis, in terms of  
band eigenstates of $H_0$ at the Fermi energy, as done in Ref.\cite{yip}.  
To leading order in $\Delta_0/\mu$, the two approaches yield identical results for the superconducting gap on the Fermi surface:   
$
\delta_{\bf n}(\bk)= \Delta \sqrt{ \tilde{k}_z^2+ (\tilde{\bf k} \cdot {\bf n})^2}% \label{dn}
$, where $\Delta = \Delta_0 \sqrt{1-m^2/\mu^2}$. %, and $\tau$'s are Pauli matrices.  
Here we have introduced a rescaled momentum $\tilde{\bf k}$ to parameterize the Fermi surface: 
\beq
\tilde{\bf k}= (vk_x,  vk_y, v_z k_z )/  \sqrt{\mu^2 - m^2}.  \label{kt}
\eeq 
$\tilde{\bf k}$ maps the ellipsoidal Fermi surface of the Hamiltonian $H_0$ to a unit sphere. 
The gap $\delta_{\bf n}(\bk)$ vanishes at two points on the equator of the Fermi surface: $\pm {\bf k}_0 = \pm k_F \hat{\bf z} \times \bf n$.  
Hence, based on this model, previous works concluded that the $E_u$ states in Cu$_x$Bi$_2$Se$_3$ have point nodes.

However, we note that $H_0$ is fully rotationally invariant around the $\hat{\bf z}$ axis. This is an artifact of the {\it first-order} $k\cdot p$ theory, which does not  include any effect of crystalline anisotropy. In reality, the crystal of Cu$_x$Bi$_2$Se$_3$ only has a discrete three-fold symmetry, and this crystalline anisotropy is responsible for pinning the direction of the $E_u$ order parameter $\bf n$.  This motivates us to take crystalline anisotropy into account and re-examine the gap structure of $E_u$ pairing. 

We find that the gap structure depends on the orientation of the order parameter $\bf n$ relative to the crystal axes: the point nodes remain present when $\bf n$ is parallel to a two-fold axes, whereas they become lifted for $\bf n$ in all other directions, resulting in a full superconducting gap. 
To illustrate this node-lifting explicitly, we add a ``hexagonal warping'' term of third order in $\bk$ to the Hamiltonian, which is allowed by the $D_{3d}$ point group symmetry of Cu$_x$Bi$_2$Se$_3$:  
\beq
H= H_0 + \lambda \sum_\bk (k_+^3 + k_-^3)   c^\dagger_\bk  \sigma_z  s_z c_\bk, \; k_\pm \equiv k_x \pm i k_y.
\label{hw}
\eeq
Here $x$ is along a two-fold axis, or equivalently, normal to a mirror plane, as shown in Fig.1. 
This hexagonal warping term arises from the spin-orbit interaction associated with crystalline anisotropy and can be regarded as the bulk counterpart of the warping term for topological insulator surface states\cite{Fu, liu}. For $\lambda \neq 0$,  the Fermi surface becomes hexagonally deformed, and more importantly, the orbital-resolved spin polarization of Bloch states in $\bf k$ space becomes modified.  % The latter will play an important role in  the gap structure of the $E_u$ superconducting state. 

\begin{figure}
\includegraphics[height=3in]{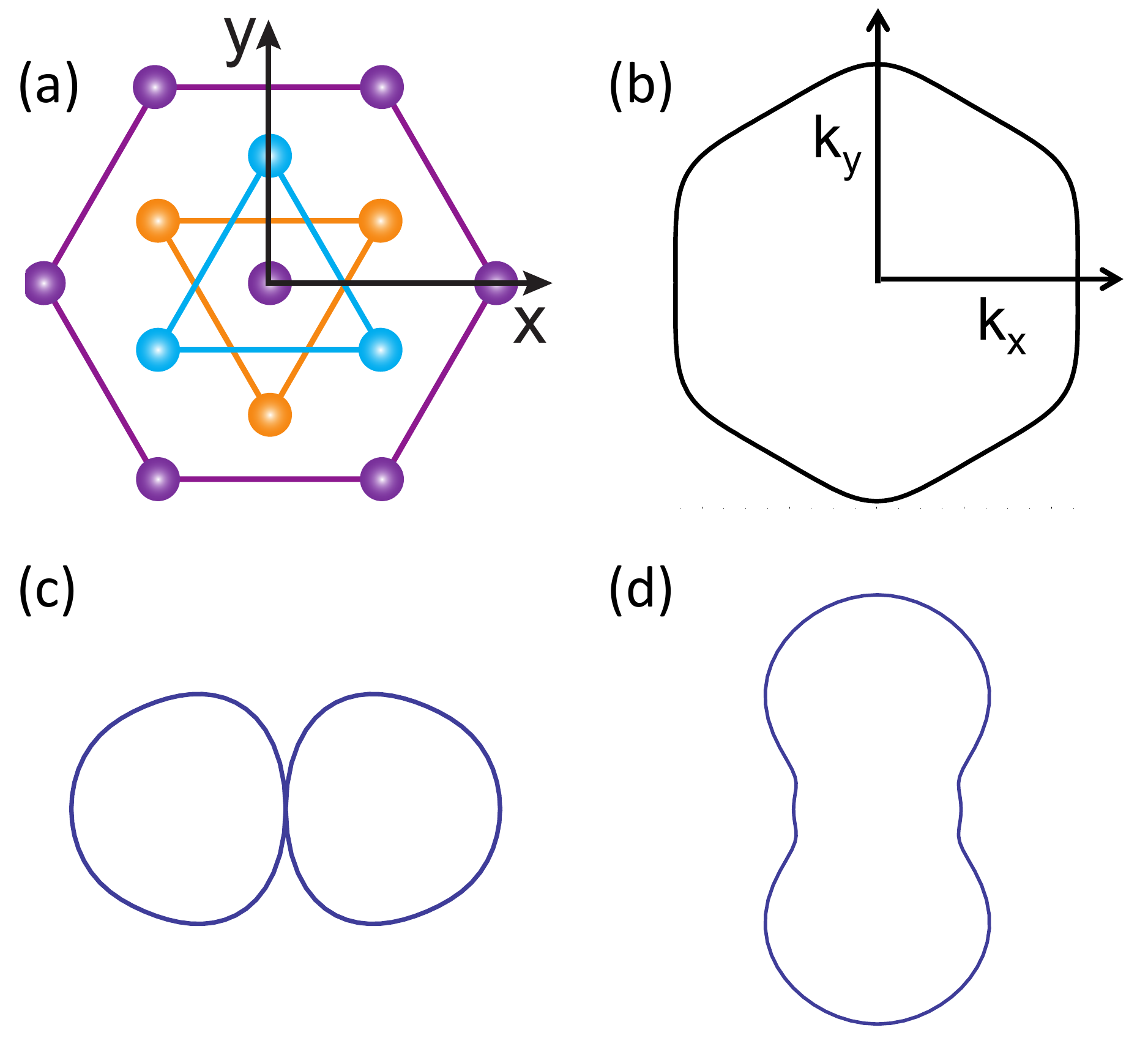}
\caption{(a) Crystal structure of Cu$_x$Bi$_2$Se$_3$ viewed from the $c$ axis. Note that the $x$ axis is normal to a mirror plane. (b) Hexagonal Fermi contour at $k_z=0$ for the Hamiltonian (\ref{hw}). (c) and (d) show the angle dependence of the anisotropic superconducting gap over the $k_z=0$ Fermi contour for the $E_u$ order parameter $V_x$ and $V_y$ respectively, defined in (\ref{pp}). The presence of nodes in (c) and the full gap in (d) are robust and model-independent.}
\end{figure}

By solving the mean-field Hamiltonian $H_{\rm sc} = H + V_{\bf n}$ with the same pair potential as before, we find the superconducting gap in the presence of hexagonal warping: 
\beq
\delta_{\bf n} (\bk) =\Delta \sqrt{1 - [\tilde{\bf k} \cdot (\hat{\bf z} \times {\bf n})]^2}, \label{gap}
\eeq    
where $\tilde{\bk}$ is still defined by Eq.(\ref{kt}), but $\bk$ now  lives on a new Fermi surface determined by  
\beq
\sqrt{m^2+ v^2 (k_x^2 + k_y^2) + \lambda^2 (k_+^3 + k_-^3)^2 + v_z^2 k_z^2} =\mu. \nonumber  
\eeq

 It is clear from (\ref{gap}) that the gap $\delta_{\bf n} (\bk)$ goes to zero only where 
$| {\bf n} \cdot (\tilde{\bk} \times \hat{\bf z}) |=1$. Importantly, we note that 
for $\lambda \neq 0$, $|\tilde{\bf k} \times \hat{\bf z}|$ is less than 1 everywhere on the warped Fermi surface, except at six corners of the hexagon on the $k_z=0$ plane (see Fig.1): $\pm {\bf k} = k_F \hat{\bf y}$ and the star of $\pm \bf k$ obtained by three-fold rotation, where $|\tilde{\bf k} \times \hat{\bf z}|=1$. As a result, 
the zero-gap condition $|\tilde{\bk} \cdot (\hat{\bf z} \times {\bf n}) |=1$ is satisfied only when the nematic director $\bf n$ is parallel to one of the three two-fold axes, such as  ${\bf n}=\pm \hat{\bf x}$. In this case, the nodes found previously remain present.  
In contrast, for ${\bf n}=(\cos \theta, \sin \theta)$ in all other directions, i.e., $\theta \neq 0, \pm \pi/3$ or $\pm 2\pi/3$,  the nodes are lifted by hexagonal warping, 
resulting in a full gap. 
   
We plot in Fig.1 the superconducting gaps over the equator of a hexagon-like Fermi surface, for two $E_u$ pairings with ${\bf n} = \hat{\bf x}$ and $\hat{\bf y}$ respectively, which are representative of the two contrasting cases. It should be said that the quantitative gap structure are model specific. For example,  
the gap anisotropy depends on the amount of warping and the microscopic  pairing interaction. Nonetheless,   
the presence of nodes for ${\bf n}= \hat{\bf x}$ and a full gap for ${\bf n}= \hat{\bf y}$, which we have explicitly shown using the model Hamiltonian (\ref{hw}) and the pair potential (\ref{pp}), are robust and model-independent properties of the $E_u$ superconducting state in Cu$_x$Bi$_2$Se$_3$, as we will show below.

Stable nodes have a deep origin in the symmetry and topology of the gap function.  
In a $T$-invariant odd-parity superconductor, a node in the gap occurs where the $d$-vector is zero. 
Importantly, we observe that when strong spin-orbit coupling is present, as in Cu$_x$Bi$_2$Se$_3$, the $d$-vector ${\vec d}(\bk)$ (whose direction depends on the choice of pseduospin basis at $\bf k$) is generically a {\it three-component} vector field in $\bf k$ space, instead of uniaxial or planar. This is simply because crystalline symmetry group alone is generally insufficient to make any component of the $d$-vector vanish {\it everywhere} in $\bf k$ space.      
Since ${\vec d}(\bk)=0$ requires satisfying three equations,  it is vanishingly improbably to find a solution on the {\it two-dimensional} Fermi surface\cite{blount}.   
 This implies that  stable nodes in T-invariant odd-parity superconductors are unlikely to occur in the presence of spin-orbit coupling, unless there is special crystal symmetry protecting their existence.  
  
An example is when there is a reflection symmetry with respect to a mirror plane, e.g., $\bf x \rightarrow -\bf x$, {\it and} the odd-parity order parameter is {\it invariant} under this reflection. In this case, ${\vec d}(k_x=0, k_y, k_z)$ and ${\vec d}(k_x=\pi/a, k_y, k_z)$ must be parallel to the normal of the mirror plane, due to its {\it pseudo}-vector nature.  
Such a  two-dimensional uniaxial $d$-vector field  on the $k_x=0, \pi/a$ plane is allowed to have lines of zeros, whose intersection with the Fermi surface will generate stable point nodes in the superconducting gap\cite{blount}.

The general argument presented above explains the gap structures of different $E_u$ states of Cu$_{x}$Bi$_2$Se$_3$ found in our model studies. 
The rotationally invariant model $H_0$ has the artifact of being symmetric with respect to any vertical plane, thus resulting in point nodes regardless of the nematic director $\bf n$. However, the crystal of Cu$_{x}$Bi$_2$Se$_3$ has only  three mirror planes that are 120 degrees apart from each other, which is correctly captured in the refined model (\ref{hw}) with hexagonal warping.     
For $\bf n$ normal to a mirror plane such as ${\bf n}=\pm \hat{\bf x}$, the corresponding order parameter $V_x$ is invariant under the reflection $\bf x \rightarrow -\bf x$; hence the nodes located on the $k_x=0$ plane are protected by this mirror symmetry.  
For $\bf n$  in other directions, however, the order parameter is not invariant under any reflection; hence nodes are absent\cite{footnote}.

To capture the important effect of crystalline anisotropy in Ginzburg-Landau theory, we must include higher-order terms in the free energy (\ref{f}), which start at the sixth order
\beq
F_6 = \kappa \left[ (\Psi^*_+\Psi_- )^3 +  (\Psi_+\Psi^*_- )^3 \right], \; \; \Psi_\pm \equiv \Psi_1 \pm i \Psi_2 
\eeq
Depending on $\kappa>0$ or $\kappa<0$, $\bf n$ is pinned either parallel or perpendicular to one of the three mirror planes, e.g., along the $\hat{\bf y}$ or $\hat{\bf x}$ axis.  
It is natural to expect that the fully gapped state with ${\bf n}=\hat{\bf y}$ has a lower free energy below $T_c$ than the nodal state with ${\bf n}=\hat{\bf x}$. 
The nematic state with ${\bf n}=\hat{\bf y}$ has two degenerate gap minima at $\pm k_F \hat{\bf x}$, and spontaneously lowers the point group symmetry from $D_{3d}$  (rhombohedral) to $C_{2h}$ (orthorhombic). This crystal symmetry breaking naturally leads to an anisotropic spin susceptbility. Importantly, the $C_{2h}$ point group in the symmetry breaking phase  has only one principal axis---the two-fold axis $\hat{\bf x}$ that lies within the $ab$-plane. It is exactly along this axis that the change in Knight shift was found to be largest in the NMR experiment\cite{zheng}. This agreement lends additional support to the $E_u$ pairing symmetry we have identified. A quantitative calculation of spin susceptibility in the anisotropic $E_u$ state depends on microscopic details, which we leave to future study.

The anisotropic $E_u$ state found here is a novel example of odd-parity pairing with a full gap.   
Among the various phases of superfluid He-3, the T-invariant B phase is isotropic, while the anistropic A phase is T-breaking. Perhaps the closest analog to  Cu$_{x}$Bi$_2$Se$_3$ is the A phase of UPt$_3$\cite{sauls}, whose order parameter is real and breaks the sixfold crystal rotational symmetry\cite{harlingen}; %, kapitulnik};   
however, this phase is known to have nodes.

{\bf Topological superconductivity}: With an odd-parity pairing symmetry and a full gap, the $E_u$ superconducting state in Cu$_x$Bi$_2$Se$_3$ satisfies all the requirements for $T$-invariant topological superconductivity stated in Ref.\cite{FuBerg}. The exact topology depends further on the nature of the Fermi surface. At low doping, the normal state has an ellipsoidal Fermi pocket centered at $\Gamma$, which under $E_u$ pairing will become a three-dimensional (3D) topological superconductor, with Majorana fermion surface states on all crystal faces. At high doping, the Fermi surface is most likely open and cylinder like, as indicated by recent photoemission\cite{ARPES} and de Haas-van Alphen measurements\cite{li1, li2}. If this is the case, the $E_u$ pairing will give rise to a quasi-two-dimensional topological superconductor, which is equivalent to stacked layers of 2D topological superconductors along the $c$ axis, correspond to $v_z=0$ in our model (\ref{hw}).  Side surfaces of this state host an even number of 2D massless Majorana fermions. The top and bottom surfaces are fully-gapped, but a step edge on these surfaces hosts 1D helical Majorana fermions. 
It has been noted\cite{ARPES} that the scenario of quasi-2D topological superconductivity may explain both the point contact and scanning tunneling spectroscopy measurements. In either 3D or quasi-2D case, more direct evidence of Majorana fermions would be desirable. 

 {\bf Experimental signatures}:
The $ab$-plane gap anisotropy of the $E_u$ pairing can be directly probed by directional dependent thermal conductivity\cite{nagai} or tunneling spectra. Here we focus on testing the $E_u$ pairing symmetry in Cu$_x$Bi$_2$Se$_3$ via the subsidiary nematic order. Symmetry dictates a linear coupling between a uniaxial strain $\epsilon_{ij}$ in the $ab$-plane and the superconducting order parameter:  
\beq
F_{\rm s} = g \left[ \frac{\epsilon_{xx} - \epsilon_{yy}}{2}(| \Psi_1|^2 - | \Psi_2|^2)  + \epsilon_{xy} (\Psi_1^* \Psi_2 + \Psi_2^* \Psi_1) \right].  \nonumber 
\eeq  
As a result of this coupling, an uniaxial strain in the $ab$-plane acts as a symmetry breaking field for the nematic order, which should be able to align the nematic director of the superconducting order parameter near $T_c$, thereby changing the pattern of the anisotropic Knight-shift. In addition, the superconducting transition temperature should increase linearly under a small uniaxial strain, independent of its direction.  The investigation of such strain-related effects on superconductivity seems within experimental reach\cite{strain} and may shed light on the pairing symmetry of Cu$_x$Bi$_2$Se$_3$. 
Furthermore, the nematic order parameter allows for half-integer disclination, around which the superconducting order parameter changes sign. 
Hence these disclinations may trap a half-integer flux quantum ($h/4e$). Finally, it would be interesting to consider whether the nematic order or other orders related to the $E_u$ pairing can emerge prior to the onset of superconductivity, similar to such phenomena in other systems\cite{nm1,nm2,nm3, intertwine}.      

I thank Guo-qing Zheng for showing the  NMR data prior to publication. This work  is supported by DOE Office of Basic Energy Sciences, Division of Materials Sciences and Engineering under award DE-SC0010526.


\begin{thebibliography}{99}
\bibitem{footnote}
We have checked that nodes are absent both on and near the Fermi surface.
\end{thebibliography}
\end{document}